\documentclass[12pt,conference]{IEEEtran}
\pdfoutput=1
\usepackage{color,graphicx}
\usepackage{amsmath}
\usepackage{enumerate}
\usepackage{syntonly}
\usepackage{multicol}
\usepackage{array}
\usepackage[tight,footnotesize]{subfigure}
\usepackage{url}
\usepackage{verbatim}
\newcommand{\degree}{\ensuremath{^\circ}}
\makeatletter

\newcommand{\Rmnum}[1]{\expandafter\@slowromancap\romannumeral #1@}
\makeatother
\begin{document}
\title{Hardware Trojan by Hot Carrier Injection}
\author{\IEEEauthorblockN{Y. Shiyanovskii, F. Wolff, C. Papachristou}
\IEEEauthorblockA{Case Western Reserve University\\
Cleveland, Ohio 44106, USA\\
\{yxs32, fxw12, cap2\}@case.edu}
\and
\IEEEauthorblockN{D. Weyer, W. Clay}
\IEEEauthorblockA{Rockwell Automation\\
\{djweyer,sclay\}@ra.rockwell.com}}
\maketitle
{\textbf Abstract.} 
This paper discusses how hot carrier injection (HCI) can be exploited
to create a trojan that will cause hardware failures.
The trojan is produced not via additional logic circuitry but
by controlled scenarios that maximize and accelerate
the HCI effect in transistors.
These scenarios range from manipulating the manufacturing
process to varying the internal voltage distribution.
This new type of trojan is difficult to test
due to its gradual hardware degradation mechanism.
This paper describes the HCI effect, detection techniques and
discusses the possibility for maliciously induced HCI trojans.

\section*{Introduction}
Due to the capital costs of building and maintaining fabrication
facilities, the number of fabs is shrinking. 
Hence, there has been a major shift in control
over the fabrication process for integrated circuits, IC.
More and more vendors outsource the fabrication process to
off shore fabrication facilities \cite{TROJANPROB08A}.
Using such facilities makes the integrated circuits vulnerable
to malicious alterations.  These alterations are more commonly
known as hardware trojans and are usually created by insertion
of additional logic circuitry \cite{HOST08a,HOST08b,HOST08c,HOST08e,HOST08g,DATE08}. The goal of these trojans ranges from
functional changes of the circuit to a complete system failure.

A new type of trojan that avoids the use of additional logic circuitry can still be induced by
exploiting the principles of semiconductor physics.
One such trojan maximizes and accelerates
the effects of Hot Carrier Injection (HCI).
HCI is a mechanism that degrades the physical characteristic of a
transistor, resulting in slower switching times or break down.
The HCI is just one of the wearing mechanisms that are responsible
for transistor degradation.
An HCI trojan can be manifested through process variations during
fabrication or changes in voltage distribution during operation.

The scaling of process technology has increased the effect of HCI on CMOS degradation.
As a result, the operational life time of CMOS devices has been 
decreasing.
Integrated circuit manufacturers have been addressing this problem
with various process and design techniques.
Fabs test for HCI during initial process characterization, but
generally do not test during wafer production.
However, due to the outsourcing shift in
manufacturing, safeguards against HCI become vulnerable.  This can 
result in undetected malicious HCI trojan.  

This paper is organized as follows:
In section two, HCI mechanism is described.
In section three, detection techniques for HCI is discussed.
Finally, in section four, different methods are explored for introducing HCI trojans.

\section*{HCI Description}

The term ``hot carrier injection'' refers to carriers
that have gained high enough kinetic energy due to intense electric
fields and are injected into the gate oxide of the CMOS device.
The degradation associated with HCI refers to the
physical breakdown and characteristic distortion of the device.
The operational conditions and the toggling frequency of
the CMOS transistor are direct contributors to the HCI rate \cite{HCI96A}.

There are four known mechanism for Hot Carrier Injection,
that describe the conditions for carriers to enter the gate oxide: SHE, CHE, DAHC, SSHE. 
The effects of HCI are more prominent in NMOS devices compared
to the PMOS devices.  It requires $3.3eV$ for electrons to
overcome the surface energy barrier at the Si--SiO$_2$ interface
and get injected into the oxide, compared to $4.6eV$ for holes.
This paper will concentrate on three HCI effects in NMOS devices. 
\begin{figure}[htp]
\centering{
\includegraphics[scale=.37]{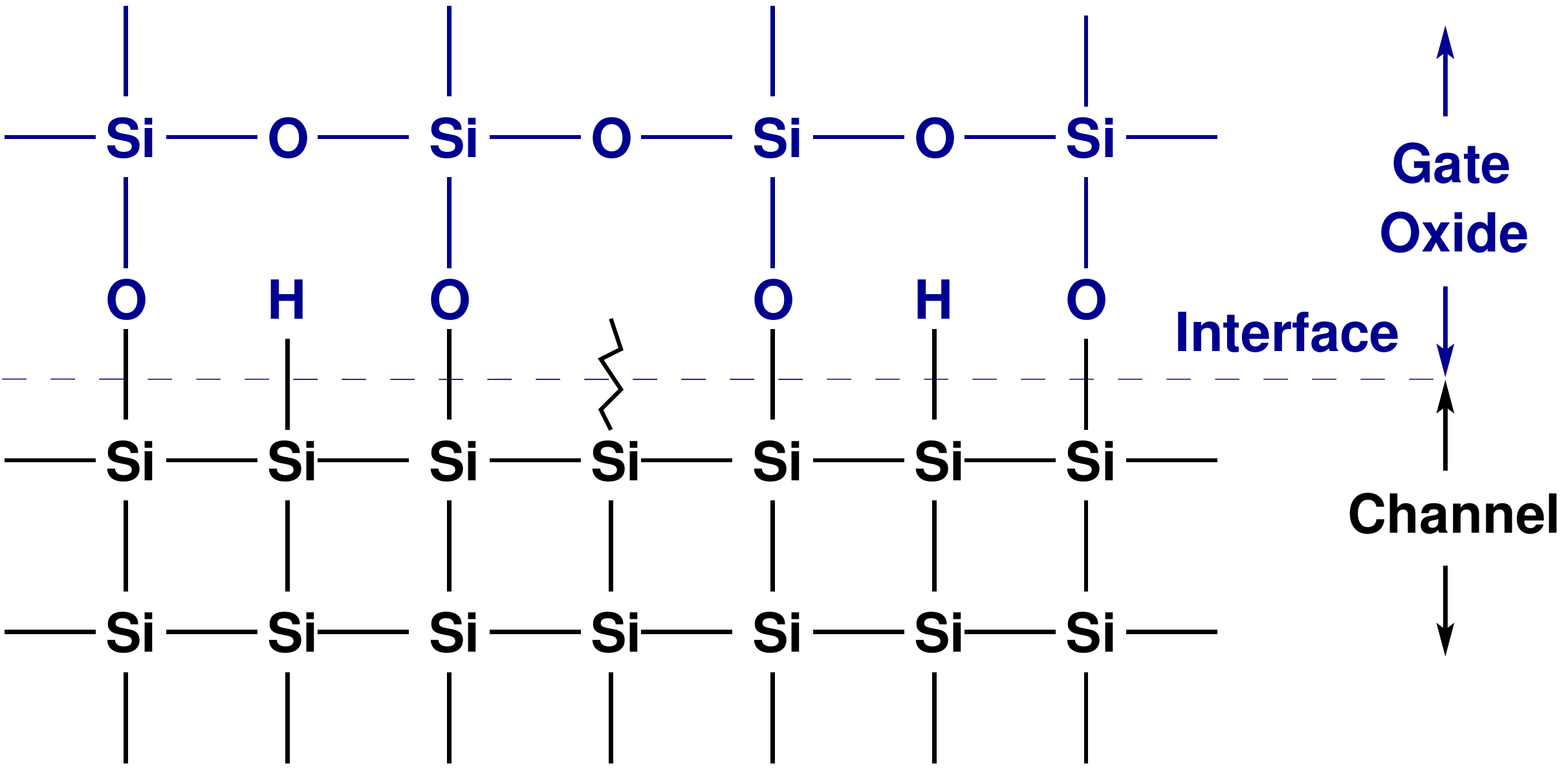}
}
\caption{Silicon Lattice interface at the gate oxide}
\label{figHCI1}
\end{figure}

\begin{figure}[htp]
\centering{
\includegraphics[scale=.3]{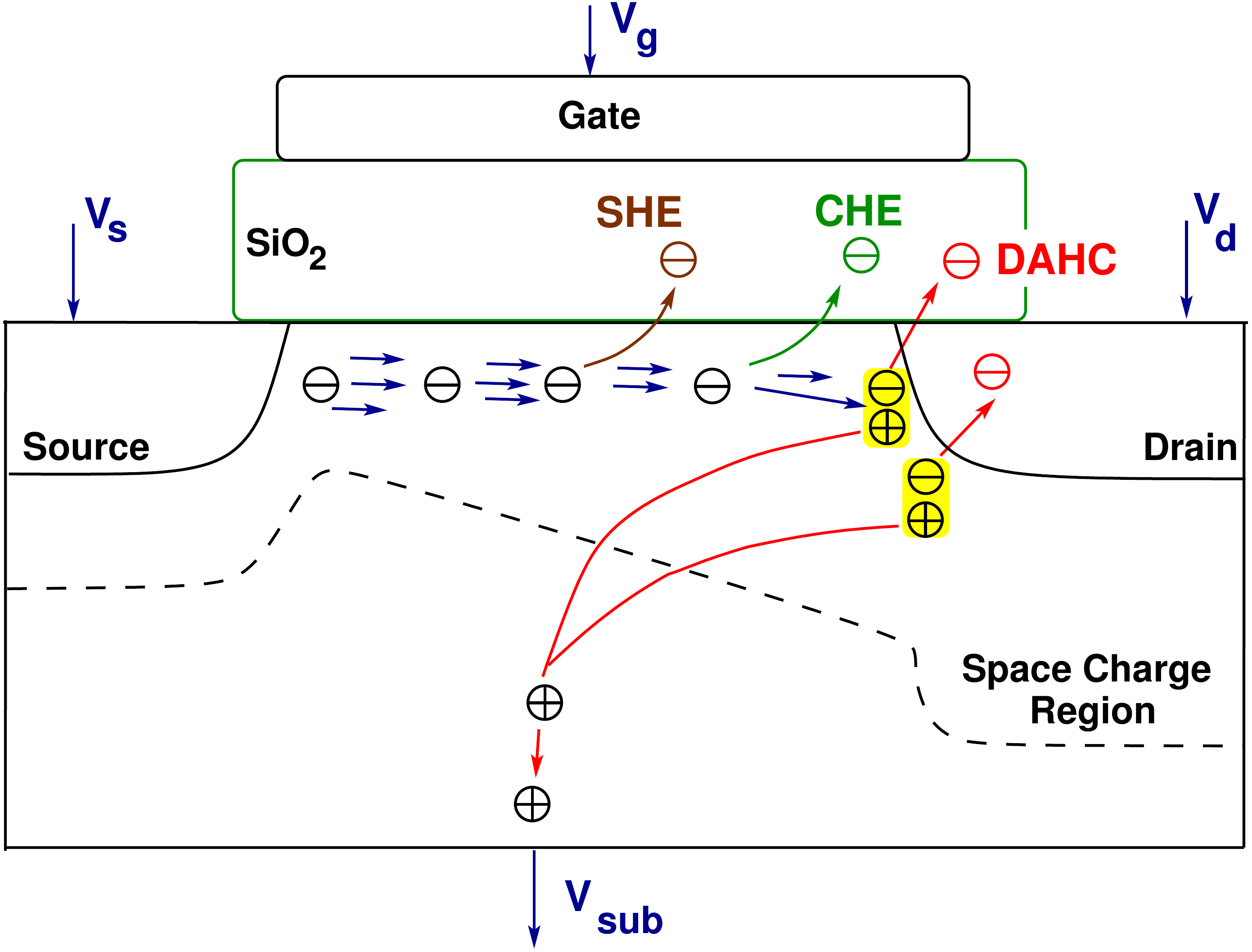}
}
\caption{HCI effects: SHE ($|V_{sub}|>>0$), CHE ($V_d=V_g$), DAHC ($V_d=2V_g$)}
\label{figHCI2}
\end{figure}

The most physically destructive HCI mechanism is Drain Avalanche
Hot Carrier (DAHC) injection.
This type of carrier injection occurs when the drain voltage,
$V_d$, is much greater then the gate voltage, $V_g$ (worse case  $V_d=2V_g$).
Such conditions create a very high electric field near the drain region.
This high electric field accelerates the carriers
into the drain depletion region.
The high rate of acceleration propels the carriers to collide
with Si lattice atoms and through impact ionization,
create displaced electron-hole pairs, shown by the yellow
region in Figure \ref{figHCI2}.
The majority of the generated holes are usually absorbed by the
substrate and thus increases the substrate current, $I_{sub}$.
Some of the generated electrons proceed to the drain and result in increased drain
current, $I_d$.  However, some of the electron - hole pairs
gain enough energy to breach the Si--SiO$_2$ interface energy
barrier, $3.3eV$ for electrons and $4.6eV$ for holes.  

Once, the carriers have passed the energy barrier
of the Si--SiO$_2$, they can either be trapped at the Si--SiO$_2$
interface, within the oxide itself or  become gate current, $I_g$.
After the bulk silicon has been cleaved  and the exposed silicon bonds have been passivated,
during the manufacturing process, Si--H bonds are formed,
shown in Figure\ref{figHCI1}. Carriers with enough energy,
$\geq 0.3eV$, can break these weak Si--H bonds and get trapped,
forming a space charge.
Over time, as more and more carriers are trapped, there is an increase in  threshold voltage, $V_t$,
and a change in  conveyed conductance, $g_m$ for NMOS transistors.
These changes in device characteristics result in decrease performance and eventual device failure.      

Substrate hot electron, SHE, injection occurs when the substrate voltage
is different than zero volts, $|V_{sub}| >> 0$.
Under this condition, an electrical field from the substrate drives
the carriers in the channel upward towards the gate,
as shown in Figure \ref{figHCI2}.
As the carriers move closer to the substrate-oxide interface,
they are influenced by high electrical field in the surface depletion region,
which accelerates the carriers and raises their kinetic energy.
If the carriers reach high enough energy, greater then $3.3eV$,
they will breach the energy barrier of the Si--SiO$_2$ interface
and behave exactly as the carriers in the DAHC scenario.  

\begin{figure}[htp]
\centering{
\includegraphics[scale=.9]{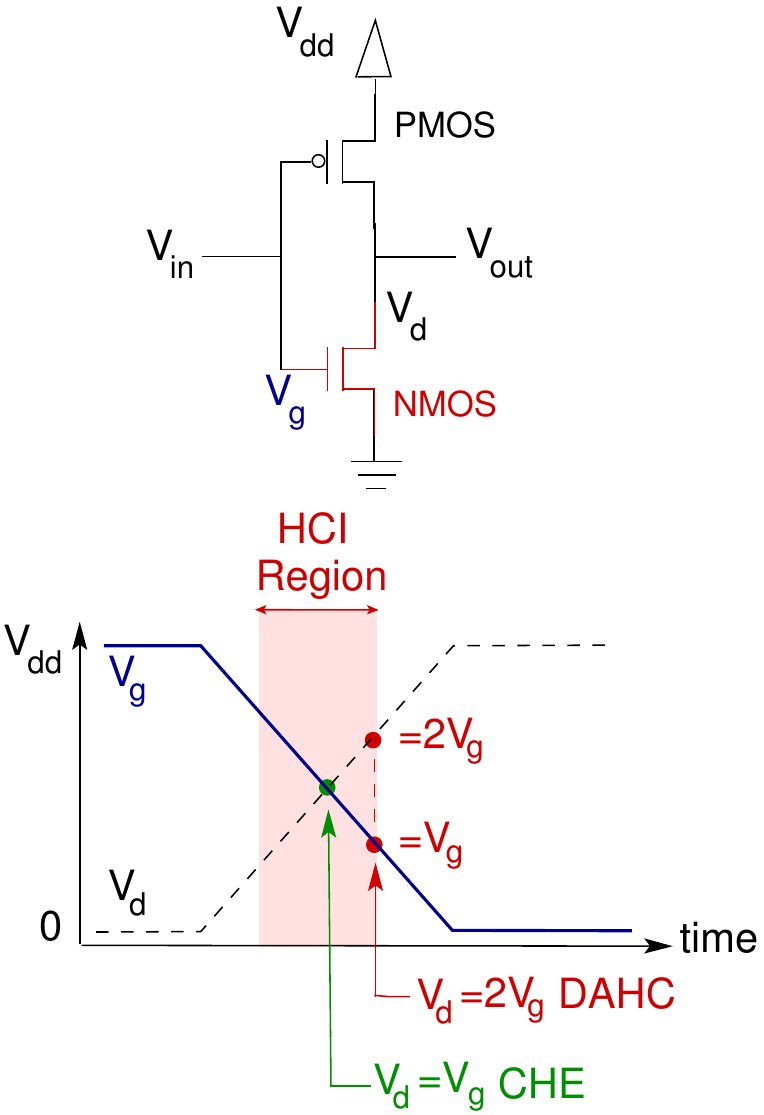}
}
\caption{Dynamic HCI effects on a cmos inverter}
\label{figINVERTER}
\end{figure}

Channel hot electron, CHE, injection occurs when the drain voltage,
$V_d$, is equal to the gate voltage, $V_g$, $V_d=V_g$,
and both voltages are higher then the source voltage, $V_s$.
Under this condition, the electric field may propel some of
the carriers into the gate oxide before they reach the drain
depletion region, as shown in Figure \ref{figHCI2}.
The behavior is exactly the same as in DAHC and SHE scenarios.
CHE and SHE do not physically damage the Si lattice atoms
during its injection mechanisms, while DAHC is very destructive
to the drain depletion region.    Figure \ref{figINVERTER} shows the DAHC and CHE effects on a 
CMOS inverter.

The static (DC) conditions for drain the, $V_d$, and substrate, $V_{sub}$,
voltages leading to the DAHC and the SHE mechanisms are
minimized through the transistor design process.  However, the CHE and DAHC
injection mechanisms do occur during the normal dynamic (AC) transistor
switching operation as shown in Figure \ref{figHCI4}.
Thus, switching frequency as well as voltage
conditions determine the rate of HCI in a transistor.  The normal design and manufacturing process are optimized for to produce a transistor operational lifetime of greater then 10 years. 

To combat the problem of gate oxide 
degradation, some manufacturing processes, $\geq 40nm$,
use nitrous oxide, $N_2O$, or nitric oxide, $NO$, during
the oxidation process \cite{NOX92A,NOX92B}.
By optimizing concentration of nitrate of the interface layer, the trap density is reduced.For a given process, there is an acceptable range of
nitrate that can be used that decreases the HCI effects and
avoids changing the performance of the CMOS device.
Concentrations of nitrogen above the acceptable range will lead to gate oxide breakdown. In summary, the HCI rate is optimized by manufacturing process and transistor design.   

\section*{HCI Detection}

There are detection techniques in place to test the influence of
HCI in an integrated circuit, IC.    HCI has many different symptoms: change in substrate
current, $I_{sub}$, change in the drain current, $I_d$, change in the
gate current, $I_g$, change in the threshold voltage, $V_t$, and
change in conveyed conductance, $g_m$.
For a packaged circuit, some of these symptoms are hard to diagnose.
Thus, most detection techniques test for known indicators ($I_{DD}$, $I_{DDQ}$, $V_t$, access time, propagation delay, ...) that are
attributed to HCI.

One detection techniques is a canary mechanism based
on delay monitoring \cite{DELAY04A,DETECT06A,DELAY07A,DELAY08A}.
This technique employs an additional circuit that compares
the propagated delay of a transistor or any number of
devices to a pre-determined stored value.
As the transistor ages, the compared values are
adjusted.
A flag is raised when the propagation delay of a transistor
is greater then expected.  This method assumes that the test circuit does not age, since it performs periodic testing and therefore not under heavy usage.
The main drawback of such method of detection is inability to distinguish the cause of the changing propagation delay,  since there are other phenomena that effect transistor delay.

Ring oscillators are another method for monitoring transistor degradation.  \cite{RING92A,RING02A}  This technique can be utilized on the wafer level, in conjunction with a canary mechanism or an external monitoring probe.   

Design techniques employ device reliability and lifetime prediction models, that result in design parameter guard-banding.  \cite{DESIGN06A,DESIGN07A,DESIGN07B} However the resulting guard-band can be inadequate to accommodate parameter variability induced by varying environmental and manufacturing conditions. Other, design techniques, such as obfuscation \cite{DESIGN08A}, hide the design but also fail in HCI detection since HCI affects the transistor and not the circuit layout. Self-calibration design\cite{SOLN07B,SOLN08Y}  techniques can be used to alter the clock or supply voltage in response to monitored transistor degradation.

\begin{figure*}[htp]
\centering{
\includegraphics[scale=.65]{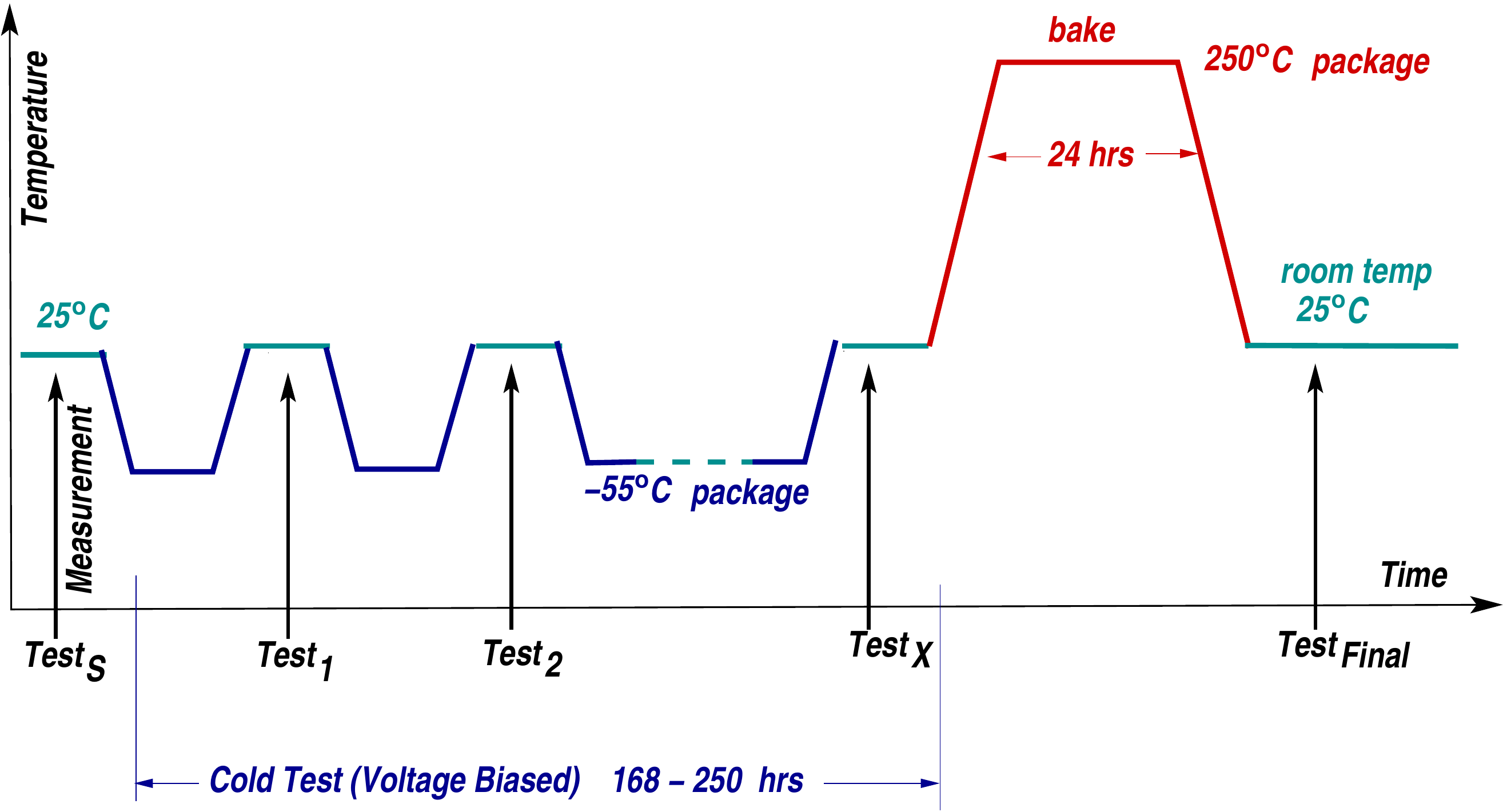}
}
\caption{Package-level diagnostic test}
\label{figHCI3}
\end{figure*}

In addition, there are wafer level \cite{TRANREL02} and package level reliability monitoring techniques, as well as parameter prognostic techniques \cite{ICOPHM08} that have been developed to detect HCI transistor degradation.  Prognostic measures can be applied at the system level to detect trends of precursors to HCI induced failure. Furthermore, there are fast wafer level reliability monitor techniques that have been recently proposed \cite{SCORED07}. 

One packaged device level detection techniques varies the environmental operating conditions to detect the presence of HCI failure mechanism.  Figure \ref{figHCI3} demonstrates the process: 
\renewcommand{\labelenumi}{\Roman{enumi}.}
\begin{enumerate}
\item Test for selected parameters to insure the test devices functionality.
\item Devices are placed in an environment is set to -$55\degree$C  and supply voltage ($V_{dd}$) is set to maximum or 24 hours.
\item Ramp devices back to room temperature, $25\degree$C.
\item Test for selected parameters.
\item If the device is still functional from step \Rmnum{5} then repeat \Rmnum{2}, \Rmnum{3}, \Rmnum{4} for a minimum of a 160 hours to a maximum of 240 hours.
\item Test for selected parameters and if there is a significant change this suggests HCI failure.
\item Perform unbiased bake at $250\degree$C for 24 hours.
\item Test at room temperature, if full or partial recovery in the monitored parameters is achieved then the failure was due to the HCI effects. 
\end{enumerate}
 The test results show the resilience of a device to the HCI effect.  Drawbacks to such a test technique are its destructive nature and the length of the test procedure.  This test technique does not provide the reason (design process, manufacturing process, etc.) for  the HCI induced failures.

\section*{HCI Trojan}
Two ways an HCI trojan could be introduced into an integrated circuit (IC), are through the design process or the manufacturing process.  The purpose is to maliciously modify transistors that will cause the IC to fail prematurely by utilizing the HCI effect.

As described in Section 2, there are manufacturing processes that minimize the effect of HCI.  Without knowledge of the circuit design, a hacker can distort the variation of nitrate concentration to create a wafer with minimal HCI resilience.  The distortion of the nitrate process will influence the lifetime of the device produced from that wafer.   The annealing time or the temperature used during the nitrate layering process can also be used to vary the resilience to HCI \cite{NOX92A},\cite{IEDM95}.

The hacker can also create ``infected'' transistors at the design level.  There are several voltage conditions for a transistor that create HCI mechanisms, they are described in section 2.  The hacker can alter the voltage supply to a particular transistor or an array of transistors to create optimal conditions for HCI.  For example, the conditions to induce DAHC, ($V_d=2V_g$), during static or dynamic transistor operation.  This method is very attractive to the hacker, because he can target a specific area of the chip to accelerate the HCI effect.

\begin{figure}[htp]
\centering{
\includegraphics[scale=.75]{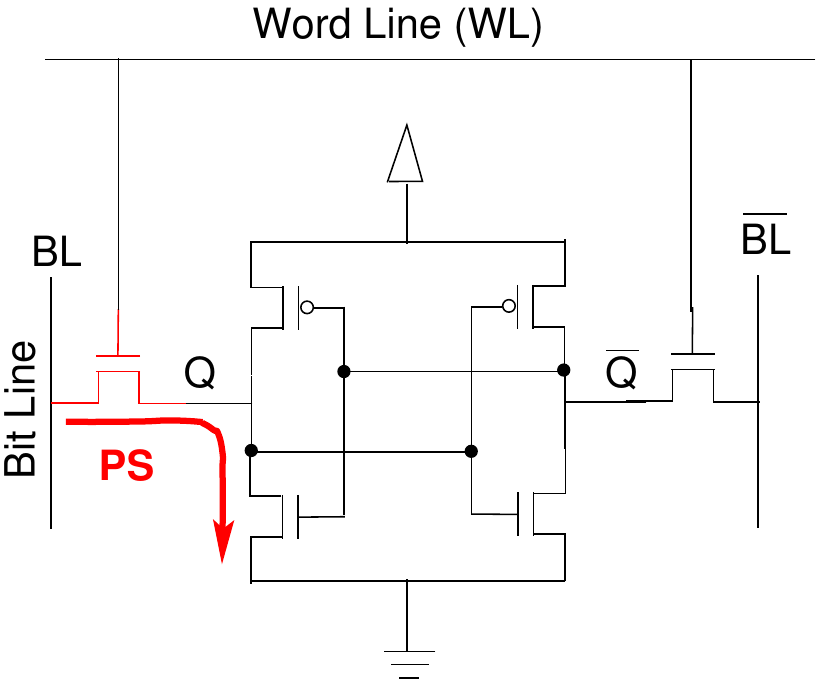}
}
\caption{HCI effects on SRAM}
\label{figHCI4}
\end{figure}

The manufacturing process variations, such as nitrate distortion, can be coupled with the design changes, such as variations in voltage distribution, for a cumulative HCI effect.  The cumulative effect will drive the infected transistors to wear even faster.  This allows the hacker to isolate the most critical transistors in order to produce maximum effect for the malicious attack with minimum circuit alteration.  

For a example in a case of a processor, the hacker can modify the cache memory module because the cache is heavy utilized and it is not protected by error correction (ECC) \cite{CACHE07A}.

In a six transistor SRAM cell which can be used in the cache, the passthrough transistor is a perfect candidate for infection by the hacker.  The passthrough transistor, seen in Figure \ref{figHCI4}, experiences heavy stress under high frequency of access.  The red arrow in Figure \ref{figHCI4} shows the current flow. An infected passthrough transistor (PS) will wear much faster then the rest of the SRAM cell.  After some time of operation, the passthrough transistor will have a very high threshold voltage, $V_t$ due to HCI.  If the gate voltage, $V_g$, is not sufficient compared to the threshold voltage, the transistor fails to function.  At this point, the passthrough transistor will not let a new value be written in during a write operation.   Such device failures can lead to catastrophic results. In addition, the hacker can apply software techniques to further accelerate the process by increasing the writing frequency of the infected cell.  The gradual degradation of the infected HCI transistors demonstrates a malicious attack, similar to  a time-bomb mechanism, that is difficult to detect.

\section*{Conclusion}
This paper discussed the concept of a new type of trojan that exploits HCI effect in transistors.  There are no detection techniques that can detect all the variations of the HCI trojans in CMOS devices.  In any post-production tests the infected transistor are operating just as well as normal transistors with no discrepancy in performance.  The danger lies in the fact that until the trojan transistor has been in operation for some time, the circuit appears as if no malicious alterations have been inserted.  The relationship between HCI effect and switching usage of a transistor, makes the trojan HCI transistors incredibly dangerous.   There is a critical demand for detection techniques that can properly identify HCI trojans.

\bibliographystyle{IEEEtran}
\bibliography{trojanhci}
\end{document}